\documentclass[aps,twocolumn,showpacs,floatfix]{revtex4}

\usepackage{bm}
\usepackage{graphicx}
\usepackage{amsmath}
\usepackage{amssymb}

\def\be{\begin{equation}}
\def\ee{\end{equation}}
\def\bea{\begin{eqnarray}}
\def\eea{\end{eqnarray}}
\def\lan{\langle}
\def\ran{\rangle}
\def\r{{\bm r}}
\def\R{{r^2}}
\def\Rea{{r^2_{\scriptscriptstyle\text{EA}}}}
\def\Rmax{{r^2_{\text{max}}}}

\begin{document}

\title{Effective Potential in Glass Forming Liquids}
\author{Antonio de Candia}
\affiliation{%
Dipartimento di Scienze Fisiche, Universit\`a di Napoli
``Federico II'' and INFM, Unit\`a di Napoli\\
Complesso Universitario di Monte Sant'Angelo,
Via Cintia, I-80126 Napoli, Italy}


\begin{abstract}
The effective potential formalism is applied to glass forming liquids,
choosing a coupling potential such that the ``order parameter'',
conjugated to the coupling strength, is the mean square displacement
of the particles from their position in the quenched reference configuration.
The potential is linear in some interval of its argument, signaling the
coexistence of two phases in the system, one with low and one with high
mean square displacement. Within this formalism, one can also compute
the free energy of metastable states and their complexity.
\end{abstract}

\maketitle

The phase space of glassy systems is believed to break up, below some
temperature $T_c$ called ``mode-coupling temperature'', into an extensive
number of components, called ``free energy valleys'' or ``metastable states'',
separated by high free energy barriers. Lowering the temperature, the number
of valleys accessible to the system becomes lower and lower, until
at some temperature $T_K$, called ``Kauzmann temperature'', it becomes
non-extensive. The existence of metastable states in the temperature
range $T_K<T<T_c$ can be evidenced by the study of a system coupled
to a quenched configuration \cite{M95,FP95,FP98,CFP97,FDPG99}.
Consider a system described by the variables $x$, and characterized by
the Hamiltonian $H(x)$. Now switch on a coupling potential
${\cal H}_\epsilon(x_0,x)=-\epsilon Q(x_0,x)$, where $x_0$ is a configuration
of the system thermalized at temperature $T$, and then kept fixed (quenched),
and $Q(x_0,x)$ is a measure of the ``similarity'' between the
configurations $x_0$ and $x$ (overlap).
The free energy per particle $f(\epsilon)$ of the coupled system
should be independent from the configuration $x_0$ in the thermodynamic
limit (self-averaging). The effective potential is defined as the
Legendre transform $v(q)=\max_{\epsilon}[f(\epsilon)+\epsilon q]$,
and corresponds to the free energy of a system constrained at a fixed overlap
$Q(x_0,x)=Nq$ from the reference configuration $x_0$, where $N$ is the number
of particles.

The effective potential $v(q)$ was computed in mean field $p$-spin models
\cite{FP95,FP98}, and in glass forming liquids using the hypernetted chain
approximation \cite{CFP97}. In both cases it was found that, besides the
thermodynamically stable minimum at $q=0$, below temperature $T_c$ a
secondary minimum appears, at high values of the overlap, signaling the
presence of metastable states. The free energy difference between the two
minima of the potential corresponds to the logarithm of the number
of free energy valleys, that is the complexity,
that vanishes at the thermodynamic glass transition $T_K$.
Below $T_K$, the thermodynamic properties of the glass can still be
evaluated by studying $m$ coupled replicas of the system, and analytically
continuing the free energy to values $m<1$ \cite{M95,MP98,CMPV99,CPV00}.
A recent review can be found in Ref.\ \cite{M02}.

It is important to point out that, in finite dimensional systems with short
range interactions, free energy barriers are finite below $T_c$
due to nucleation effects (that is activated or hopping processes),
and diverge only at $T_K$. For temperatures $T_K<T<T_c$,
the time needed to escape from a free energy valley is large but finite.
Moreover the coexistence of different phases,
implies that the effective potential
has to be a convex function of its argument.
This means that $v(q)$ cannot have a secondary minimum, and the dynamical
transition $T_c$ cannot be sharply defined.

In this paper, I will apply the effective potential formalism to
simple liquids, choosing a particularly convenient coupling potential.
The potential is such that the ``order parameter'', conjugated to the coupling
strength $\epsilon$, is the mean square displacement
(m.s.d.)\ of the particles from their position in the
quenched reference configuration \cite{nota1}.
Let us consider a liquid composed by $N$ monoatomic particles of mass $m$ 
enclosed in a cubic box of side $L$, interacting with
a potential energy $U(\r)$, where $\r$ is a $3N$-dimensional vector
representing the positions of the particles,
with periodic boundary conditions. The equilibrium free energy
per particle, at temperature $T$, is given by (setting $k_B=1$)
\be
f_{\text{eq}}(T)=-\frac{T}{N}\ln%
\!\!\!\int\limits_{\scriptscriptstyle 0<r_i<L}
\!\!\!\frac{d^{\scriptscriptstyle 3N}\!{\bf r}}{N!\lambda^{3N}}\,\,
e^{-\beta U(\r)},
\label{free}
\ee
where the integral is done over an hypercubic box of volume $V^N$,
with $V=L^3$, and $\lambda=\hbar\sqrt{2\pi/mT}$ is the thermal wavelength.
Let us now consider a (quenched) configuration $\r_0$
thermalized at temperature $T$, and define the coupling potential
\be
{\cal H}_\epsilon(\r_0,\r)=-T\ln\left[(\epsilon/\epsilon_0)^{\frac{3N}{2}}
\sum_{P,{\cal T}}e^{-\beta\epsilon |\r-P{\cal T}\r_0|^2}\right],
\label{Heps}
\ee
where $\epsilon\ge 0$ is the coupling strength, and
$\epsilon_0=\pi T(\rho/e)^{2/3}$, where $\rho=N/V$ is the density
of the liquid. The operator
${\cal T}$ runs over all the discrete translations of step $L$
in the $3N$-dimensional space, and the operator $P$ runs
over all the permutations of the particles. The configuration
$P{\cal T}\r_0$ runs therefore over all the configurations
``thermodynamically identical'' to $\r_0$.
It may be said that configuration $\r_0$ is ``partially annealed'',
because those identical configurations are effectively summed over in the 
partition function.
When $\epsilon\to 0$ the potential ${\cal H}_\epsilon(\r_0,\r)$ vanishes,
as can be seen doing the substitution
\be
\sum_{P,{\cal T}}e^{-\beta\epsilon|\r-P{\cal T}\r_0|^2}
\>\>\>\raisebox{-6pt}{%
$\stackrel{\displaystyle\longrightarrow}{\scriptstyle\epsilon\to 0}$}\>\>\>\>
\frac{N!}{V^N}\int%
\!d^{\scriptscriptstyle 3N}\!{\bf r}_0\,e^{-\beta\epsilon|\r-\r_0|^2}.
\label{sumint}
\ee
It can be seen, analogously, that
$\frac{\partial {\cal H}_\epsilon}{\partial\epsilon}(\r_0,\r)$ vanishes as well
when $\epsilon\to 0$.
On the other hand, when $\epsilon>0$,
the potential favors the configurations that,
apart from a permutation of the particles, or a discrete translation 
of step $L$, are ``near'' to the configuration $\r_0$.
We can now evaluate the coupled free energy as
\bea
&&f(\epsilon)=-\frac{T}{N}\ln%
\!\!\!\int\limits_{\scriptscriptstyle 0<r_i<L}
\!\!\!\frac{d^{\scriptscriptstyle 3N}\!{\bf r}}{N!\lambda^{3N}}\,\,
e^{-\beta [U(\r)+{\cal H}_\epsilon(\r_0,\r)]}
\nonumber\\
&&=-\frac{T}{N}\ln\left\{(\epsilon/\epsilon_0)^{\frac{3N}{2}}
\!\!\!\!\!\!\!\!\!\int\limits_{\scriptscriptstyle -\infty<r_i<\infty}
\!\!\!\!\!\!\!\frac{d^{\scriptscriptstyle 3N}\!{\bf r}}{\lambda^{3N}}\,\,
e^{-\beta U(\r)-\beta\epsilon|\r-\r_0|^2}\right\},
\label{feps}
\eea
where in the last member $U(\r)$ has been extended by periodicity
outside the domain $0<r_i<L$. The free energy $f(\epsilon)$
can be also written as
\be
f(\epsilon)=-\frac{T}{N}\ln\,\left\{(\epsilon/\epsilon_0)^{\frac{3N}{2}}
\!\!\int\limits_0^\infty
\!\!d\R\,\, e^{-\beta N[\,v(\R)+\epsilon \R\,]}\right\},
\ee
where the effective potential $v(\R)$ is defined by
\be
v(\R)=-\frac{T}{N}\ln
\!\!\!\!\!\!\!\!\!\int\limits_{\scriptscriptstyle -\infty<r_i<\infty}
\!\!\!\!\!\!\!\!\frac{d^{\scriptscriptstyle 3N}\!{\bf r}}{\lambda^{3N}}\,
\delta(\R-\tfrac{1}{N}|\r-\r_0|^2)
\,e^{-\beta U(\r)}.
\label{effpot}
\ee
The derivative of the effective potential with respect to $\R$
can be expressed as
\be
\frac{\partial\,v(\R)}{\partial\R}=\frac{3T}{2\R}
\left[\vartheta(\R)-1\right],
\label{vder}
\ee
where $\vartheta(\R)=\beta\lan(\r-\r_0)\cdot\nabla U\ran_\R/3N$,
and the average $\lan\cdots\ran_\R$
is done at fixed m.s.d.\ $\R$ from the configuration $\r_0$,
with the Boltzmann weight $e^{-\beta U(\r)}$.
From Eq.\ (\ref{effpot}), being $\r_0$ thermalized
at temperature $T$, it can be derived that $\lim_{\R\to 0}[v(\R)
+(3T/2)\ln(2e\pi\R/3\lambda^2)]=u_{\text{eq}}(T)$,
where $u_{\text{eq}}(T)$ is the mean potential energy per particle.
The effective potential
can then be computed by integration of $\vartheta(\R)$,
\be
v(\R)=u_{\text{eq}}(T)+\!\frac{3T}{2}\!\left[
\ln\!\left(\frac{3\lambda^2}{2e\pi\R}\right)\!
+\!\!\int\limits_{0}^{\,\,\R}\!\vartheta(\R)\,d\!\ln\R\!\right].
\label{thetaint}
\ee

It is reasonable to believe that $f(\epsilon)$, $v(\R)$ and $\vartheta(\R)$
are self-averaging quantities, that is in the thermodynamic limit
they do not depend on the configuration $\r_0$, but only on the temperature
at which $\r_0$ is thermalized.
The potential $v(\R)$ can be interpreted as the free energy of a system
constrained to be at m.s.d.\ $\R$ from the reference configuration $\r_0$.
Actually, this is true only if $\R$ is not too large. Otherwise,
many ``thermodynamically identical'' configurations,
differing only by a permutation of the particles
or a discrete translation of step $L$, will be contained in the
hypersphere $\tfrac{1}{N}|\r-\r_0|^2=\R$, and some ``Gibbs
correction factor'' will be needed in the configurational integral in
Eq.\ (\ref{effpot}).
The effective potential $v(\R)$ and the free energy $f(\epsilon)$ are 
related by the Legendre transform
\be
v(\R)=\max_{\epsilon}\left[\,f(\epsilon)
+\frac{3T}{2}\ln\,(\epsilon/\epsilon_0)-\epsilon\R\right].
\label{legendre}
\ee
Note that $f(\epsilon)+\frac{3T}{2}\ln\,(\epsilon/\epsilon_0)$
is a convex function. Being $\lim_{\epsilon\to 0}f(\epsilon)=f_{\text{eq}}(T)$,
and $\lim_{\epsilon\to 0}\frac{\partial f}{\partial\epsilon}(\epsilon)=0$
\cite{nota2}, from Eq.\ (\ref{legendre}) one then finds
\be
\lim_{\R\to\infty}\left[v(\R)+\frac{3T}{2}\ln(\R/\Rmax)\right]
=f_{\text{eq}}(T),
\label{vlarge}
\ee
where $\Rmax=3/(2e^{1/3}\pi\rho^{2/3})$.
This equation has a simple interpretation: when $\R$ is large,
the hypersphere $\tfrac{1}{N}|\r_0-\r|^2=\R$
contains every configuration of the liquid (modulo a permutation
or a translation of step $L$) a number 
of times given by the ratio between the volume
of the hypersphere and the volume $V^N/N!$ of the ``physical''
phase space, that is $(\R/\Rmax)^{3N/2}$.
This is the ``Gibbs correction factor'', by which
the configurational integral in Eq.\ (\ref{effpot}) should be divided,
when $\R>\Rmax$, to obtain the equilibrium free energy.
Therefore, the effective potential $v(\R)$ represents the constrained
free energy only when $\R<\Rmax$. On the other hand, when $\R>\Rmax$, 
the liquid is essentially not constrained, and its free energy is
$f_{\text{eq}}(T)$.
Putting together Eq.\ (\ref{thetaint}) and (\ref{vlarge}), the 
equilibrium free energy can be computed, obtaining
\be
f_{\text{eq}}(T)=u_{\text{eq}}(T)+T\!\left[%
\ln(\rho\lambda^3/e)\!
+\!\frac{3}{2}\!\!\int\limits_{0}^{\infty}\!\!
\vartheta(\R)\,d\!\ln\R\!\right]\!.
\label{freeq}
\ee

In a ``mean field'' picture, it may happen that the effective potential $v(\R)$
has a metastable minimum at some ``Edwards-Anderson'' m.s.d.\ $\Rea$.
As it was mentioned before, in a finite dimensional system with short range
interactions, the free energy barriers remains finite when the number of
particles goes to infinity, due to nucleation effects. Phase coexistence
then implies that $v(\R)$ is a convex function \cite{nota3}.
Nevertheless, at temperatures near or below the mode-coupling
temperature $T_c$, if we let the system evolve from the configuration $\r_0$,
it will remain trapped for a very long time at some m.s.d.\
$\Rea$ from $\r_0$, that corresponds to the {\em plateau} in $\lan\R(t)\ran$
observed in the free evolution.
If the time needed to escape from the metastable state is longer than the
time needed to thermalize inside it, one can effectively define the free
energy of the metastable state as the constrained free energy $v(\Rea)$.
The complexity, defined as $\Sigma=\frac{1}{N}\ln{\cal N}$,
where ${\cal N}$ is the number of metastable states, can then
be computed as the difference between
the constrained free energy $v(\Rea)$, and the equilibrium one
$f_{\text{eq}}(T)$, divided by the temperature. We are supposing here
that the ``constrained potential energy'' $\lan U(\r)\ran_\Rea/N$
coincides with the equilibrium one $u_{\text{eq}}(T)$, which is in 
principle assured only if $\Rea$ is a real minimum of $v(\R)$, 
so that metastable states have infinite lifetime.

\begin{figure}
\includegraphics[width=7.5cm]{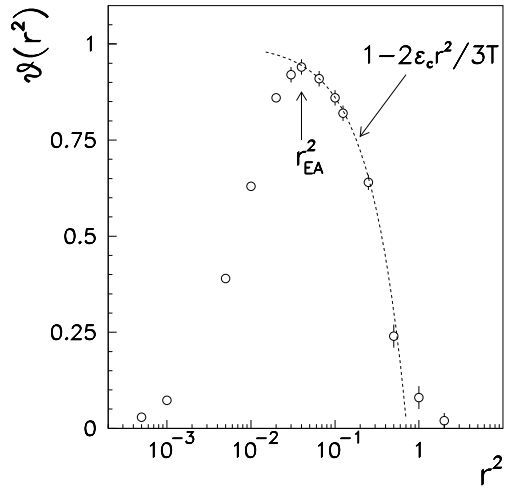}
\caption{Open circles: function $\vartheta(\R)$ in LJBM
for temperature $T=0.5$. The Edwards-Anderson m.s.d.\ is chosen
(somewhat arbitrarily) to be $\Rea=0.04$, that corresponds to the maximum
of $\vartheta(\R)$, and is also close to the {\em plateau} in
$\lan\R(t)\ran$ observed in the free evolution.
The interval in which points are well fitted by the dashed line is possibly
characterized by phase coexistence. The critical coupling,
corresponding to the (putative) first order transition,
is $\epsilon_c=1.05\pm0.10$ for this temperature.}
\label{fig1}
\end{figure}

When temperature is low enough, a first order transition line in the
$\epsilon-T$ plane is expected to appear. When this happens, there
will be an interval in $\R$, corresponding to the coexistence of two
different phases in the system, where the effective potential
will be a linear function of $\R$, namely $v(\R)=v_0-\epsilon_c\R$,
where $\epsilon_c$ is the critical value of $\epsilon$ for that temperature.
In terms of $\vartheta(\R)$, in the coexistence interval it will be
$\vartheta(\R)=1-\tfrac{2{\textstyle\epsilon_c}}{3T}\,\R$.

This formalism can be applied to a hard sphere liquid as well.
In that case, evaluating $\lan(\r-\r_0)\cdot\nabla U\ran_\R$
as a temporal average, and being the time integral of the force $-\nabla U$
equal to the momentum variation $\Delta{\bm p}$,
the function $\vartheta(\R)$ can be computed as
\be
\vartheta(\R)=\lim_{\Delta t\to\infty}
\frac{\beta}{3N\Delta t}\sum_i(\r_0-\r_i)\cdot\Delta{\bm p}_i\,
\raisebox{-4pt}{\Big|}_{\R}
\ee
where the sum is over all the collisions having place in the interval of time
$\Delta t$, $\r_i$ and $\Delta{\bm p}_i$ are the
coordinates and momentum variations at the time of the $i$-th collision,
and the liquid is constrained to move at fixed m.s.d.\ $\R$ 
from the configuration $\r_0$.

\begin{figure}
\includegraphics[width=7.5cm]{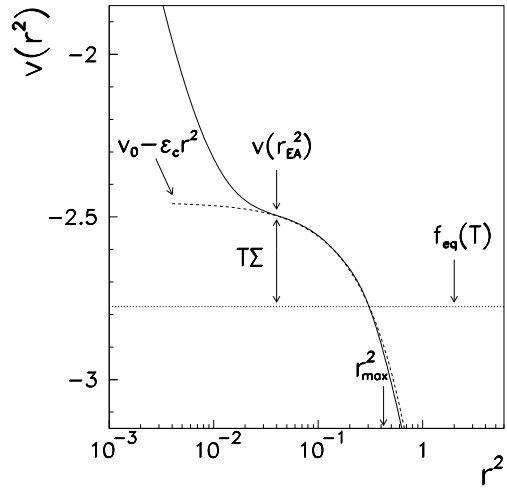}
\caption{Solid line: effective potential $v(\R)$ in LJBM
for temperature $T=0.5$. The potential is computed by Eq.\ (\ref{thetaint}),
integrating a function that interpolates the measured values of
$\vartheta(\R)$, and setting $\hbar=m=1$.
The dotted line corresponds to $f_{\text{eq}}(T)$,
computed by Eq.\ (\ref{freeq}) corrected for mixtures (see text).
As in Fig.\ \ref{fig1}, the interval fitted by the dashed line is possibly
characterized by phase coexistence. The difference
$[v(\Rea)-f_{\text{eq}}(T)]/T$ gives %
the complexity, that is $\Sigma=0.55\pm 0.03$ for this temperature.}
\label{fig2}
\end{figure}

I have computed the function $\vartheta(\R)$ in a well known binary mixture
of particles interacting by a Lennard-Jones potential (LJBM) \cite{KA95},
namely a mixture of 80\% A
particles and 20\% B particles, with
pair potential $v_{\alpha\beta}(r)=4\,\epsilon_{\alpha\beta}\,
[(\sigma_{\alpha\beta}/r)^{12}-(\sigma_{\alpha\beta}/r)^6]$,
where $\epsilon_{AA}=1$,
$\sigma_{AA}=1$, $\epsilon_{AB}=1.5$, $\sigma_{AB}=0.8$,
$\epsilon_{BB}=0.5$, and $\sigma_{BB}=0.88$.
I simulate the system at density $\rho=(0.94)^{-3}$, where the mode-coupling
temperature is $T_c=0.435$ \cite{KA95}.
In the case of mixtures, in previous equations one has to substitute $N!$
with $N_A!N_B!$. Moreover $\epsilon_0$
has to be multiplied by the factor $(x_A^{x_A} x_B^{x_B})^{2/3}$,
where $x_A$ and $x_B$ are the fractions of A and B particles,
and $\Rmax$
has to be divided by it.
Therefore for this system one has $\Rmax=0.4220$.
Finally, the free energy in Eq.\ (\ref{freeq}) gets an extra contribution
$T(x_A\ln x_A+x_B\ln x_B)$.

The number of particles simulated is $N=1000$, in a box of side $L=9.4$,
with pair potential truncated at a distance $2.5\,\sigma_{\alpha\beta}$
and shifted. First of all, a configuration $\r_0$ is thermalized at
temperature $T$, and kept fixed. Then, the system is cloned, copying
configuration $\r_0$ into new variables $\r$. Subsequently, configuration
$\r$ is allowed to evolve, until the m.s.d.\ $\tfrac{1}{N}|\r-\r_0|^2$
is equal to $\R$. At this point, while $\r_0$ is still kept fixed, $\r$ is
simulated constraining the m.s.d. to remain constant, by means of the
{\small RATTLE} algorithm \cite{AT87}.
It is then necessary to ``thermalize'' configuration $\r$,
by letting it evolve onto the hypersphere for a sufficient time.
Finally, a time average of $\beta[(\r-\r_0)\cdot\nabla U]/3N$ is computed
to evaluate $\vartheta(\R)$. This procedure can be repeated many times,
with different and independent configurations $\r_0$, to average
``over the disorder'', and evaluate the errors as standard deviations.

\begin{figure}
\includegraphics[width=7.5cm]{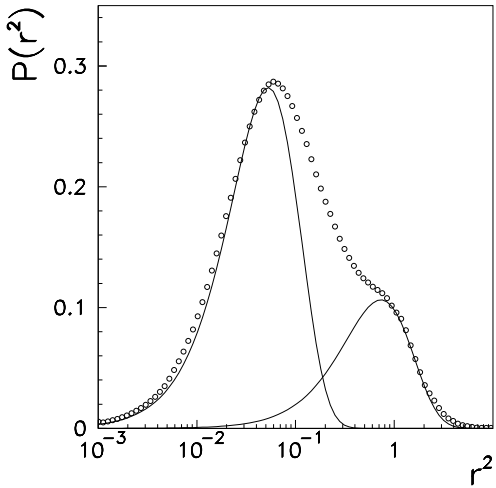}
\caption{Open circles: distribution of the square displacements
of the particles, when the m.s.d.\ is constrained to be $\lan r^2\ran=0.25$.
Solid lines: two Gaussian distributions, with m.s.d.\
$0.053$ and $0.73$, and weights $0.61$ and $0.23$.}
\label{fig3}
\end{figure}

In Fig.\ \ref{fig1}, the function $\vartheta(\R)$ for temperature $T=0.5$
is plotted. It shows a maximum at $\R=0.04$, that is very close to
the {\em plateau} observed in $\lan\R(t)\ran$, when the system
is allowed to evolve freely. This can be then identified with the
Edwards-Anderson m.s.d. $\Rea$.
The interval where $\vartheta(\R)$ is well fitted by the
function $1-\tfrac{2{\textstyle\epsilon_c}}{3T}\,\R$ corresponds to the
coexistence of two phases in the system. The measured values of
$\vartheta(\R)$ were interpolated by a smooth function, that was then
integrated to obtain the effective potential $v(\R)$ by Eq.\ (\ref{thetaint}),
which is plotted in Fig.\ \ref{fig2}. Here, the interval of phase
coexistence is signaled by the fit with the function $v_0-\epsilon_c\R$.
Note that the m.s.d.\ $\Rmax$ marks the crossover between the constrained
liquid for $\R<\Rmax$, whose free energy is $v(\R)$, and the
``unconstrained'' liquid for $\R>\Rmax$, whose
free energy is $f_{\text{eq}}(T)$. The complexity for this temperature
is equal to $\Sigma=0.55\pm 0.03$, a value compatible with the one 
measured in the inherent structure formalism
within the ``harmonic approximation'' \cite{SKT99}.

When the system is homogeneous, the distribution 
of the square displacements of the particles is expected to be 
Gaussian, $P(r^2)\propto r^2\exp(-3r^2/2\lan r^2\ran)$
(here we denote by $r^2$ the square displacement of a single particle,
and by $\lan r^2\ran$ the m.s.d.)
On the other hand, when two phases coexist in the system, we
expect that the distribution is given by the superposition of two
Gaussians, plus a contribution from the interface. This is shown in
Fig.\ \ref{fig3}, where the distribution measured in the system constrained
at m.s.d.\ $\lan r^2\ran=0.25$ is plotted. On the average, a fraction
$0.61$ of the particles belong to the ``bound phase'', with m.s.d.\
$0.053$, and a fraction $0.23$ to the ``unbound phase'', with m.s.d.\
$0.73$. The remaining fraction of particles makes up the interface between
the two phases. This ``non-Gaussian heterogeneity'' shows up also in
the free evolution of the system \cite{HET97}.

In conclusion, I have applied the effective potential formalism to glass
forming liquids, revealing the presence of a first order transition
in the $\epsilon-T$ plane, between two phases characterized by low and
high mean square displacement with respect to a quenched
reference configuration. It would be interesting to understand how much
of this picture is relevant to the dynamical behavior of the system.
Indeed, near or below the mode-coupling temperature $T_c$,
the relaxation times may be connected to the free energy
barrier that must be overcome to nucleate a ``unbound bubble'' inside
the sea of bound particles.

Acknowledgments --
It is a pleasure to thank A. Coniglio for many fruitful discussions.
This work was partially supported by INFM-PRA (HOP),
MURST-PRIN-2003, and MIUR-FIRB-2002.


\begin{thebibliography}{99}
\bibitem{nota1} A similar coupling potential was already considered
in Ref.\ \cite{CMPV99}, in order to compute the free energy of
metastable states.
\bibitem{nota2} This can be derived noting that
$\frac{\partial f}{\partial\epsilon}(\epsilon)$ is the
thermal average of
$\frac{\partial {\cal H}_\epsilon}{\partial\epsilon}(\r_0,\r)$,
that goes to zero when $\epsilon\to 0$.
\bibitem{nota3} Note however that $v(\R)$ has not to be convex
when $\R>\Rmax$, because in that region it does not represent
the constrained free energy.
\end{thebibliography}
\end{document}